\documentclass[final,5p,times, sort&compress, fleqn,twocolumn]{elsarticle}
\journal{Journal of Magnetism and Magnetic Materials}
\usepackage{amsmath}
\usepackage{bm}

\begin{document}

\begin{frontmatter}

  \title{
    Precession dynamics of a small magnet with non-Markovian damping: Theoretical proposal for an experiment to determine the correlation time\tnoteref{grant}
  }

  \author{Hiroshi Imamura}
  \ead{h-imamura@aist.go.jp}

  \author{Hiroko Arai}
  \ead{arai-h@aist.go.jp}

  \author{Rie Matsumoto}

  \author{Toshiki Yamaji}

  \author{Hiroshi Tsukahara\tnoteref{keke}}

  \address{National Institute of Advanced Industrial Science and Technology (AIST), Tsukuba, Ibaraki 305-8568, Japan}

  \tnotetext[kek]{Permanent address: High Energy Accelerator Research Organization (KEK), Institute of Materials Structure Science (IMSS), Tsukuba, Ibaraki 305-0801, Japan}

  \tnotetext[grant]{This work is partly supported by JSPS KAKENHI Grant Numbers JP19H01108 and JP18H03787.}

  \begin{abstract}
    Recent advances in experimental techniques have made it possible to manipulate and measure the magnetization dynamics on the femtosecond time scale which is the same order as the correlation time of the bath degrees of freedom. In the equations of motion of magnetization, the correlation of the bath is represented by the non-Markovian damping. For development of the science and technologies based on the ultrafast magnetization dynamics it is important to understand how the magnetization dynamics depend on the correlation time. It is also important to determine the correlation time experimentally. Here we study the precession dynamics of a small magnet with the non-Markovian damping. Extending the theoretical analysis of Miyazaki and Seki [J. Chem. Phys. {\bf 108}, 7052 (1998)] we obtain analytical expressions of the precession angular velocity and the effective damping constant for any values of the correlation time under assumption of small Gilbert damping constant. We also propose a possible experiment for determination of the correlation time.
  \end{abstract}

\end{frontmatter}

\section{Introduction}
Dynamics of magnetization is the combination of precession and damping. The precession is caused by the torque due to the internal and external magnetic fields. Typical time scale of the precession around the external field and the anisotropy field is nanosecond. The damping is caused by the coupling with the bath degrees of freedom such as conduction electrons and phonons. The typical time scale of the relaxation of conduction electrons and phonons is picosecond or sub-picosecond which is much faster than precession. In typical experimental situations such as ferromagnetic resonance and magnetization process, the time correlation of the bath degrees of freedom can be neglected and the magnetization dynamics is well represented by the Landau-Lifshitz-Gilbert (LLG) equation with the Markovian damping term\cite{landau1935,gilbert2004,brown1963}.

Recent advances in experimental techniques such as femtosecond laser pulse and time-resolved magneto-optical Kerr effect measurement have made it possible to manipulate and measure magnetization dynamics on the femtosecond time scale\cite{beaurepaire1996,stanciu2007,zhang2009,bigot2009,kirilyuk2010a,bigot2013,walowski2016a,quessab2018}.
In 1996, Beaurepaire et al. observed the femtosecond laser pulse induced sub-picosecond demagnetization of a Ni thin film\cite{beaurepaire1996}, which opens the field of ultrafast magnetization dynamics. The all-optical switching of magnetization in a ferrimagnetic GdFeCo alloy was demonstrated by Stanciu et al. using a 40 femtosecond circularly polarized laser pulse\cite{stanciu2007}. The helicity-dependent laser-induced domain wall motion in Co/Pt multilayer thin films was reported by Quessab et al.\cite{quessab2018}.

To understand the physics behind the ultrafast magnetization dynamics it is necessary to take into account the time correlation of bath in the equations of motion of magnetization. The first attempt was done by Kawabata in 1972\cite{kawabata1972}. He derived the Bloch equation and the Fokker-Planck equation for a classical spin interacting with the surroundings based on the Nakajima-Zwanzig-Mori formalism\cite{nakajima1958,zwanzig1960,mori1965}. In 1998, Miyazaki and Seki constructed a theory for the Brownian motion of a classical spin and derived the integro-differential form of the generalized Langevin equation with non-Markovian damping\cite{miyazaki1998}. They also showed that the generalized Langevin equation reduces to the LLG equation with modified parameters in a certain limit. Atxitia et al. applied the theory of Miyazaki and Seki to the atomistic model simulations and showed that materials with smaller correlation time demagnetized faster\cite{atxitia2009}.

Despite the experimental and theoretical progresses to date little attention has been paid to how to determine the correlation time experimentally. For development of the science and technologies based on the ultrafast magnetization dynamics it is important to determine the correlation time experimentally as well as to understand how the magnetization dynamics depend on the correlation time.

In this paper the precession dynamics of a small magnet with non-Markovian damping is theoretically studied based on the macrospin model. The magnet is assumed to have a uniaxial anisotropy and to be subjected to an external magnetic field parallel to the magnetization easy axis. The non-Markovianity enhances the precession angular velocity and reduces the damping. Assuming that the Gilbert damping constant is much smaller than unity, the analytical expressions of the precession angular velocity and the effective damping constant are derived for any values of the correlation time by extending the analysis of Miyazaki and Seki\cite{miyazaki1998}. We also propose a possible experiment for determination of the correlation time. The correlation time can be determined by analyzing the external field at which the enhancement of the precession angular velocity is maximized.

The paper is organized as follows. Section \ref{sec:model} explains the theoretical model and the equations of motion. Section \ref{sec:hkzero} gives the numerical and theoretical analysis of the precession dynamics in the absence of an anisotropy field. The effect of the anisotropy field is discussed in Sec. \ref{sec:anis}. A possible experiment for determination of the correlation time is proposed in Sec. \ref{sec:possible}. The results are summarized in Sec. \ref{sec:summary}.

\section{Theoretical model}
\label{sec:model}
We calculate the magnetization dynamics in a small magnet with a uniaxial anisotropy under an external magnetic field based on the macrospin model. The magnetization easy axis is taken to be $z$-axis and the magnetic field is applied in the positive $z$-direction. In terms of the magnetization unit vector, $\bm{m}=(m_{x},m_{y},m_{z})$, the energy density is given by
\begin{align}
  E
  =
  K(1-m_{z}^{2})-\mu_{0}\, M_{s}\, H\, m_{z},
\end{align}
where $K$ is the effective anisotropy constant including the crystalline, interfacial, and shape anisotropies. $\mu_{0}$ is the vacuum permeability, $M_{s}$ is the saturation magnetization, $H$ is the external magnetic field. The effective field is obtained as
\begin{align}
  \bm{H}_{\rm eff}
  = (H_{k}\, m_{z} + H)\bm{e}_{z},
\end{align}
where $\bm{e}_{z}$ is the unit vector in the positive $z$ direction and $H_{k} = 2 K/(\mu_{0}\,M_{s})$ is the effective anisotropy field.

The magnetization precesses around the effective field with damping. The energy and angular momentum are absorbed by the bath degrees of freedom such as conduction electrons and phonons until the magnetization becomes parallel to the effective field. The equations of motion of $\bm{m}$ coupled with the bath is given by the Langevin equation with the stochastic field representing the bath degrees of freedom. If the time scale of the bath is much smaller than the precession frequency the stochastic field can be treated as the Wiener process\cite{gardiner2009} as shown by Brown\cite{brown1963}.

Since we are interested in the ultrafast magnetization dynamics of which time scale is the same order as the correlation time of the bath degrees of freedom, the stochastic field should be treated as the Ornstein-Uhlenbeck process\cite{uhlenbeck1930,gardiner2009}. As shown by Miyazaki and Seki \cite{miyazaki1998} the equations of motion of $\bm{m}$ takes the following integro-differential form:
\begin{align}
  \label{eq:gL}
  \dot{\bm{m}}
   & =
  -\gamma\,
  \bm{m}\times\left(\bm{H}_{\rm eff} + \bm{r}\right)
  +
  \alpha\,
  \bm{m} \times \int_{-\infty}^{t} \nu(t-t')\, \dot{\bm{m}}(t')\, dt',
\end{align}
where $\gamma$ is the gyromagnetic ratio, $\alpha$ is the Gilbert damping constant, and $\bm{r}$ is the stochastic field.
The first term represents the precession around the sum of the effective field and the stochastic field, and the second term represents the non-Markovian damping.
The memory function in the non-Markovian damping term is defined as
\begin{align}
  \nu(t-t') = \frac{1}{\tau_{c}}\exp\left(-\frac{|t-t'|}{\tau_{c}}\right),
\end{align}
where $\tau_{c}$ is the correlation time of the bath degrees of freedom.
The stochastic field, $\bm{r}$, satisfies $\langle r_{i}(t)\rangle = 0$ and
\begin{align}
  \label{eq:rr_correlation}
  \langle r_{j}(t)\, r_{k}(t')\rangle = \frac{\mu}{2}\,\delta_{j,k}\,\nu(t-t'),
\end{align}
where $\langle \ \rangle$ represents the statistical mean, and
\begin{align}
  \mu=\frac{2\alpha\, k_{\rm B}\,T}{\gamma\,M_{s}\, V}.
\end{align}
The subscripts $j$ and $k$ stand for $x$, $y$, or $z$, $k_{\rm B}$ is the Boltzmann constant, $T$ is the temperature, $V$ is the volume of the magnet, and $\delta_{j,k}$ is Kronecker's delta.
The LLG equation with the Markovian damping derived by Brown \cite{brown1963} is reproduced in the limit of $\tau_{c}\to 0$ because $\lim_{\tau_{c}\to 0}\nu(t-t') = 2\delta(t-t')$, where $\delta(t-t')$ is Dirac's delta function.
Equation \eqref{eq:gL} is equivalent to the following set of the first order differential equations,
\begin{align}
  \label{eq:gL_diff1}
   & \dot{\bm{m}}
  =
  -\gamma \bm{m}\times \left[ \bm{H}_{\rm eff} + \delta\bm{H} \right]
  \\
  \label{eq:gL_diff2}
   & \dot{\delta \bm{H}}
  =
  -\frac{1}{\tau_{c}}\delta\bm{H} - \frac{\alpha}{\tau_{c}^{2}} \bm{m}
  -\frac{\gamma}{\tau_{c}}\bm{R},
\end{align}
where $\bm{R}$ represents the stochastic field due to thermal agitation.
Equations \eqref{eq:gL_diff1}, \eqref{eq:gL_diff2} are used for numerical simulations.
The stochastic field, $\bm{R}$, satisfies $\langle R_{j}(t)\rangle = 0$ and
\begin{align}
  \label{eq:RR_correlation}
   & \langle R_{j}(t)\, R_{k}(t')\rangle = \mu\,\delta_{j,k}\,\delta(t-t').
\end{align}

\section{Precession dynamics in the absence of an anisotropy field}
\label{sec:hkzero}
In this section the precession dynamics in the absence of an anisotropy field, i.e. $H_{k} = 0$, is considered. The initial direction of magnetization is assumed to be $\bm{m}=(1,0,0)$.  The numerical simulation shows that the non-Markovian damping enhances the precession angular velocity and reduces the damping. The numerical results are theoretically analyzed assuming that $\alpha \ll 1$. The analytical expressions of the precession angular velocity and the effective damping constant are obtained. The case with $H_{k}\neq 0$ will be discussed in Sec. \ref{sec:anis}.

\subsection{numerical simulation results}
\begin{figure}[t]
  \centerline{\includegraphics [width=\columnwidth] {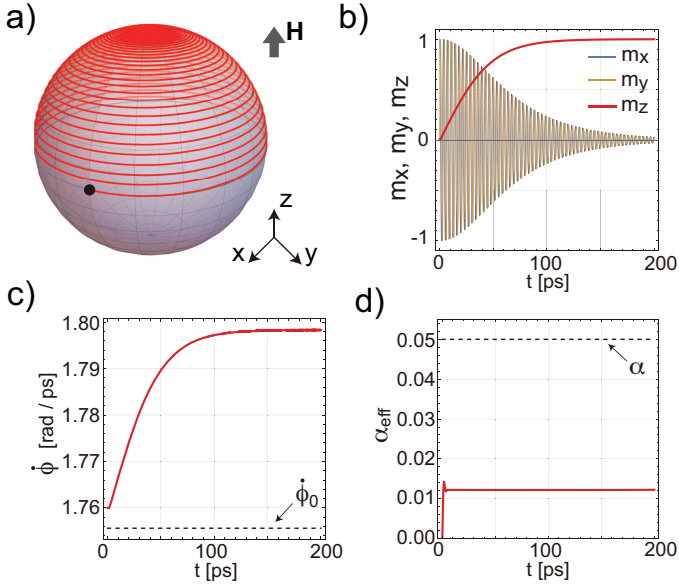}}
  \caption{
    \label{fig1}
    (a) Trajectory of $\bm{m}$ on a unit sphere. The external field of $H$ = 10 T is applied in the positive $z$ direction. The initial direction is assumed to be $\bm{m}=(1,0,0)$ as indicated by the filled circle. The other parameters are $\tau_{c} = 1$ ps, and $\alpha = 0.05$.
    (b) Temporal evolution of $m_{x}$,  $m_{y}$,  $m_{z}$.
    (c) Temporal evolution of the precession angular velocity, $\dot{\phi}$. The solid red curve shows the simulation result. The dotted black line indicates the result of the Markovian LLG equation, i.e. $\dot{\phi}_{0}=\gamma H/(1+\alpha^{2})$.
    (d) Temporal evolution of the effective damping constant, $\alpha_{\rm eff}$. The solid red curve shows the simulation result. The dotted black line indicates $\alpha=0.05$.
  }
\end{figure}
We numerically solve Eqs. \eqref{eq:gL_diff1}, \eqref{eq:gL_diff2} for $H = 10$ T, $\alpha = 0.05$, and $\tau_{c}=1$ ps. The temperature is assumed to be low enough to set $\bm{R}=0$ in Eq. \eqref{eq:gL_diff2}.
Figure \ref{fig1}(a) shows the trajectory of $\bm{m}$ on a unit sphere. The initial direction is indicated by the filled circle. The plot of the temporal evolutions of $m_{x}$, $m_{y}$, and $m_{z}$ are shown in Fig. \ref{fig1}(b). The magnetization relaxes to the positive $z$ direction with precessing around the external field. The results are quite similar to that of the Markovian LLG equation, which implies that the non-Markovianity in damping causes renormalization of the gyromagnetic ratio and the Gilbert damping constant in the Markovian LLG equation.

The renormalized value of the gyromagnetic ratio can be observed as a variation of the precession angular velocity, $\dot{\phi}$, where the polar and azimuthal angles are defined as $\bm{m} = (\sin\theta\cos\phi,\sin\theta\sin\phi,\cos\theta)$.  Figure \ref{fig1}(c) shows that temporal evolution of $\dot{\phi}$ (solid red) together with the precession angular velocity without non-Makovianity, $\dot{\phi}_{0}=\gamma H/(1+\alpha^{2})$, (dotted black). The precession angular velocity increases with increase of time and saturates to a certain value around 1.798. The shape of the time dependence of $\dot{\phi}$ is quite similar to that of $m_{z}$ shown in Fig. \ref{fig1}(b), which suggests that the non-Markovian damping acts as an effective anisotropy field in the precession dynamics.

The renormalization of the Gilbert damping constant can be observed as a variation of the temporal evolution of the polar angle, $\dot{\theta}$. Rearranging the LLG equation for $\dot{\theta}$, the effective damping constant can be defined as
\begin{align}
  \label{eq:alpha_eff}
  \alpha_{\rm eff} = -\dot{\theta}/(\gamma H \sin\theta).
\end{align}
In Fig. \ref{fig1}(d) $\alpha_{\rm eff}$ is shown by the red solid curve as a function of time. The effective damping constant is reduced to about one-fifth of the original value of $\alpha=0.05$ (dotted black). Contrary to $\dot{\phi}$, $\alpha_{\rm eff}$ does not show clear correlation with the dynamics of $\bm{m}$. During the precession, $\alpha_{\rm eff}$ is kept almost constant.

The enhancement of the precession angular velocity and the reduction of the Gilbert damping constant due to the non-Markovian damping will be explained by deriving the effective LLG equation that is valid up to the first order of $\alpha$ in the next subsection.


\subsection{Theoretical analysis }
\label{sec:approx}
Since the Gilbert damping constant, $\alpha$, of a conventional magnet is of the order of $0.01 \sim 0.1$, it is natural to take the first order of $\alpha$ to derive the effective equations of motion for $\bm{m}$. The other parameters related to the motion of $\bm{m}$ are $\gamma$, $H$, and $\tau_{c}$. Multiplying these parameters we can obtain the dimensionless parameter, $\xi=\gamma H \tau_{c}$, which represents the increment of the precession angle during the correlation time.

In the case of $\xi<1$ Miyazaki and Seki dereived the effective LLG equation using time derivative series expansion\cite{miyazaki1998}. We first briefly review their analysis. Then we derive the effective LLG equation for $\xi>1$ using the time-integral series expansion and show that the effective LLG equation has the same form for both $\xi<1$ and $\xi>1$. Therefore, it is natural to assume that the derived effective LLG equation is valid for any values of $\xi$ including $\xi=1$.

\subsubsection{Brief review of Miyazaki and Seki's derivation of the effective LLG equation for $\xi<1$ }
In Ref. \citenum{miyazaki1998}, Miyazaki and Seki derived the effective LLG equation with renormalized parameters using the time derivative series expansion. Similar analysis of the LLG equation was also done by Shul in the study of the damping due to strain\cite{suhl1998,suhl2007}. The following is the brief summary of the derivation.

Successive application of the integration by parts
using $ \nu(t-t')=\tau_{c}\,\left[d\nu(t-t')/dt'\right]$ gives the following time derivative series expansion:
\begin{align}
  \label{eq:nm_int_series}
  \int_{-\infty}^{t}\nu(t-t')\, \dot{\bm{m}}(t')\, dt'
  =\sum_{n=1}^{\infty}
  (-\tau_{c})^{n-1}\frac{d^{n}\bm{m}}{dt^{n}}.
\end{align}
Then the non-Markovian damping term in Eq. \eqref{eq:gL} is expressed as
\begin{align}
  \label{eq:nm_series}
  \alpha\,\sum_{n=1}^{\infty}
  (-\tau_{c})^{n-1}
  \left(\bm{m}\times\frac{d^{n}\bm{m}}{dt^{n}}\right).
\end{align}
The first derivative, $n=1$, is given by
\begin{align}
  \label{eq:mdot1}
  \dot{\bm{m}}
  =-\gamma H \bm{m}\times\bm{e}_{z} + O(\alpha),
\end{align}
where $O$ is the Bachmann–Landau symbol.
For $n=2$, substitution of Eq. \eqref{eq:mdot1} into the time derivative of Eq. \eqref{eq:mdot1} gives
\begin{align}
  \ddot{\bm{m}}
  =\left(-\gamma\, H\right)^{2}\left(\bm{m}\times\bm{e}_{z}\right)\times\bm{e}_{z} + O(\alpha).
\end{align}
The $n$-th order time derivative is obtained by using the same algebra as
\begin{align}
  \frac{d^{n}}{dt^{n}}
  \bm{m}
  =\left(-\gamma\, H\right)^{n}\,
  \left[\left(\bm{m}\times\bm{e}_{z}\right)\times\bm{e}_{z}\dots\right] + O(\alpha),
\end{align}
where $\bm{e}_{z}$ appears $n$ times.
Expanding the vector products we obtain for even order time derivatives
\begin{align}
  \label{eq:dmdt_even}
   & \frac{d^{2n}\bm{m}}{dt^{2n}}
  =
  (-1)^{n}
  (\gamma\,H)^{2n}
  \left[
    \bm{m}
    -
    m_{z}
    \bm{e}_{z}
    \right]
  +O(\alpha),
\end{align}
and for odd order time derivatives
\begin{align}
  \label{eq:dmdt_odd}
   & \frac{d^{2n+1} \bm{m}}{dt^{2n+1}}
  =
  (-1)^{n}
  (\gamma\,H)^{2n}\,
  \dot{\bm{m}}
  +O(\alpha).
\end{align}
Substituting Eqs. \eqref{eq:dmdt_even} and \eqref{eq:dmdt_odd} into Eq. \eqref{eq:nm_series} the non-Markovian damping term is expressed as
\begin{align}
  \label{eq:contrib_nonMarkov}
  -\left(\sum_{n=1}^{\infty}\gamma_{2n}\right)\bm{m}\times\bm{e}_{z}
  +\left(\sum_{n=0}^{\infty}\alpha_{2n+1}\right)\,\bm{m}\times\dot{\bm{m}},
\end{align}
where
\begin{align}
  \label{eq:gamma2n}
   & \gamma_{2n}
  =\alpha\,\gamma\,H\,m_{z}\,(-1)^{n-1}
  \xi^{2n-1}
  \\
   & \alpha_{2n+1}
  =
  \alpha\,
  (-1)^{n} \xi^{2n}.
\end{align}
The sums in Eq. \eqref{eq:contrib_nonMarkov} converge for $\xi<1$.
Introducing
\begin{align}
  \label{eq:gamma_tilde}
   &
  \tilde{\gamma}
  =
  \gamma \left(1+ \alpha\,m_{z}
  \frac{\xi}{1+\xi^{2}}\right)
  \\
  \label{eq:alpha_tilde}
   & \tilde{\alpha} = \frac{\alpha}{1+\xi^{2}},
\end{align}
Eq. \eqref{eq:gL} can be expressed as the following effective LLG equation with renormalized gyromagnetic ratio, $\tilde{\gamma}$, and damping constant, $\tilde{\alpha}$:
\begin{align}
  \label{eq:effectiveLLG}
  \dot{\bm{m}}=-\tilde{\gamma}\,\bm{m}\times
  \left(\bm{H}+\bm{r}\right) + \tilde{\alpha}\,\bm{m}\times\dot{\bm{m}} + O(\alpha^{2}).
\end{align}

\subsubsection{Derivation of the effective LLG equation for $\xi>1$}
For $\xi > 1$ we expand Eq. \eqref{eq:gL} in power series of $1/\xi$ using the time integral series expansion approach.
Using the integration by parts with $d\nu(t-t')/dt'=\nu(t-t')/\tau_{c}$
the integral part of the non-Markovian damping can be written as
\begin{align}
   & \int_{-\infty}^{t}\nu(t-t')\, \dot{\bm{m}}(t')\, dt'
  =
  \frac{1}{\tau_{c}}
  \int_{-\infty}^{t}\dot{\bm{m}}(t')\, dt'\nonumber       \\
   & \hspace{1em}
  -
  \frac{1}{\tau_{c}}
  \int_{-\infty}^{t}\nu(t-t')
  \left[\int_{-\infty}^{t'}\dot{\bm{m}}(t'')\, dt''\right]
  \, dt'.
\end{align}
Successive application of the integration by parts gives
\begin{align}
  \label{eq:nu_J}
  \int_{-\infty}^{t}\nu(t-t')\, \dot{\bm{m}}(t')\, dt'
  =
  -\sum_{n=1}^{\infty}
  \left(-\frac{1}{\tau_{c}}\right)^{n}
  \bm{J}_{n},
\end{align}
where $J_{n}$ is the $n$th order multiple integral defined as
\begin{align}
  \label{eq:jn}
  \bm{J}_{n}=
  \int_{-\infty}^{t}
  \int_{-\infty}^{t_{1}}
  \cdots
  \int_{-\infty}^{t_{n-1}}
  \dot{\bm{m}}(t_{n})\,
  dt_{n}
  \cdots
  dt_{2}
  dt_{1}.
\end{align}
From Eq. \eqref{eq:dmdt_odd}, on the other hand, $\dot{\bm{m}}$ is expressed as
\begin{align}
  \label{eq:dmdt_odd2}
  \dot{\bm{m}}
  =\frac{1}{(-1)^{n}(\gamma\,H)^{2n}}
  \frac{d^{2n}}{dt^{2n}}\dot{\bm{m}}
  + O(\alpha).
\end{align}
Substituting Eq. \eqref{eq:dmdt_odd2} into Eq. \eqref{eq:jn} the multiple integrals are calculated as
\begin{align}
   & \bm{J}_{2n}=\frac{1}{(-1)^{n}(\gamma\,H)^{2n}}\dot{\bm{m}}
  \\
   & \bm{J}_{2n-1}=\frac{1}{(-1)^{n}(\gamma\,H)^{2n}}\ddot{\bm{m}}.
\end{align}
Then Eq. \eqref{eq:nu_J} becomes
\begin{align}
  \label{eq:dterm4}
   & \int_{-\infty}^{t}\nu(t-t')\, \dot{\bm{m}}(t')\, dt'
  =
  \sum_{n=1}^{\infty}
  \frac{1}{(-1)^{n-1}\xi^{2n}}\dot{\bm{m}}
  \nonumber                                               \\
   & \hspace{1em}
  +
  \sum_{n=1}^{\infty}
  \frac{\tau_{c}}{(-1)^{n}\xi^{2n}}\ddot{\bm{m}}.
\end{align}
Substituting Eq. \eqref{eq:dterm4} into the second term of Eq. \eqref{eq:gL} the non-Markovian damping term is expressed as
\begin{align}
  \label{eq:nm_int1}
  \alpha
  \sum_{n=1}^{\infty}
  \frac{1}{(-1)^{n-1}\xi^{2n}}\bm{m}\times
  \left[
    \dot{\bm{m}}
    +
    \tau_{c} \ddot{\bm{m}}
    \right] + O(\alpha^{2}).
\end{align}
From Eq. \eqref{eq:dmdt_even} $\ddot{\bm{m}}$ is expressed as
\begin{align}
  \label{eq:mdot}
  \ddot{\bm{m}} =
  (-1)
  (\gamma\,H)^{2}
  \left[
    \bm{m}
    -
    m_{z}
    \bm{e}_{z}
    \right].
\end{align}
Substituting Eqs. \eqref{eq:nm_int1} and \eqref{eq:mdot} into Eq. \eqref{eq:gL} we obtain
\begin{align}
  \label{eq:gle_hh2}
  \dot{\bm{m}}
   & =-\gamma H
  \sum_{n=1}^{\infty}
  \left[
    1 + \frac{\alpha m_{z}}{(-1)^{n-1}\xi^{2n-1}}
    \right]
  \bm{m}\times\bm{e}_{z}
  - \gamma \bm{m}\times\bm{r}
  \nonumber     \\
   &
  + \alpha\,
  \sum_{n=1}^{\infty}
  \frac{1}{(-1)^{n-1}\xi^{2n}}\bm{m}\times\dot{\bm{m}}
  + O(\alpha^{2}).
\end{align}
The sums converge for $\xi>1$, and the effective LLG equation for $\xi>1$ has the same form as $\xi<1$, i.e. Eq. \eqref{eq:effectiveLLG}. Since the effective LLG equation has the same form for both $\xi< 1$ and $\xi>1$, it is natural to Eq. \eqref{eq:effectiveLLG} is valid for any values of $\xi$.

As pointed out by Miyazaki and Seki, and independently by Suhl the effect of the non-Markovian damping on the precession can be regarded as the renormalization of the effective field \cite{miyazaki1998,suhl1998,suhl2007}. Equation \eqref{eq:effectiveLLG} can be expressed as
\begin{align}
  \label{eq:fictHk}
  \dot{\bm{m}}
   & =
  -\gamma \mathbf{m}\times
  \left(
  H
  +  \frac{\alpha\, H\, \xi}{1+\xi^{2}} m_{z}
  \right)  \bm{e}_{z}  -\gamma \mathbf{m}\times\bm{r}
  \nonumber       \\
   & \hspace{1em}
  +
  \tilde{\alpha}\,\bm{m}\times\dot{\bm{m}} + O(\alpha^{2}).
\end{align}
The second term in the bracket represents the fictitious uniaxial anisotropy field originated from the non-Markovian damping. The fictitious anisotropy field increases with increase of $\xi$ for $\xi<1$ and takes the maximum value of $\alpha\, H\, m_{z}/2$ at $\xi=1$, i.e. $\gamma\,H\,\tau_{c}=1$. For $\xi>1$ the fictitious anisotropy field decreases with increase of $\xi$ and vanishes in the limit of $\xi\to\infty$ because the non-Markovian damping term vanishes in the limit of $\tau_{c}\to\infty$. The precession angular velocity, $\dot{\phi}$, is expected to have the same $\xi$ dependence as the fictitious anisotropy field and to have the same temporal evolution as $m_{z}$ as shown in Figs. \ref{fig1}(b) and \ref{fig1}(c).

\subsubsection{The Correlation time dependence of the precession angular velocity,  and effective damping constant}
Equation \eqref{eq:gamma_tilde} tells us that up to the first order of $\alpha$ the precession angular velocity can be approximated as
\begin{align}
  \label{eq:phidot_approx}
  \dot{\phi} \simeq \tilde{\gamma}H=\gamma H  \left[1+ \alpha\,m_{z}
  \frac{\gamma\,H\,\tau_{c}}{1+(\gamma\,H\,\tau_{c})^{2}}\right],
\end{align}
where the second term in the square bracket represents the enhancement due to the fictitious anisotropy field.

\begin{figure}[t]
  \centerline{\includegraphics [width=\columnwidth] {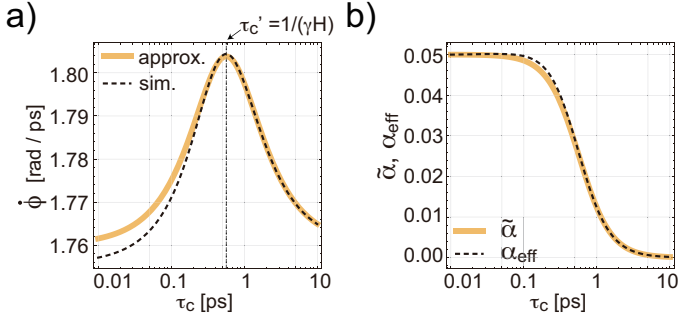}}
  \caption{
  \label{fig2}
  (a) The correlation time, $\tau_{c}$, dependence of the precession angular velocity, $\delta\dot{\phi}$, at $\theta=5^{\circ}$ for $H=10$ T. The solid yellow curve shows the approximation result, $\tilde{\gamma} H$. The dotted black curve shows the simulation results obtained by numerically solving Eqs. \eqref{eq:gL_diff1} and \eqref{eq:gL_diff2}. The thin vertical dotted line indicates the critical value of the correlation time, $\tau_{c}^{\prime}=1/(\gamma H)$.
  (b) $\tau_{c}$ dependence of $\tilde{\alpha}$ (solid yellow) and $\alpha_{\rm eff}$ (dotted black). The parameters and the other symbols are the same as panel (a).
  }
\end{figure}

In Fig. \ref{fig2}(a) the approximation result of Eq. \eqref{eq:phidot_approx} at $\theta=5^{\circ}$ where $\dot{\phi}$ is almost saturated is plotted as a function of $\tau_{c}$ by the solid yellow curve. The external field and the Gilbert damping constant are assumed to be $H=10$ T and $\alpha=0.05$, respectively. The corresponding simulation results obtained by numerically solving Eqs. \eqref{eq:gL_diff1} and \eqref{eq:gL_diff2} are shown by the dotted black curve. Both curves agree well with each other because $\alpha$ is as small as 0.05. The precession angular velocity is maximized at the critical value of the correlation time $\tau_{c}^{\prime} = 1/(\gamma H)$.

Figure \ref{fig2}(b) shows the $\tau_{c}$ dependence of $\tilde{\alpha}$ (solid yellow) and $\alpha_{\rm eff}$ (dotted black) for the same parameters as panel (a). Both curves agree well with each other and are monotonic decreasing functions of $\tau_{c}$. They vanish in the limit of $\tau_{c}\to\infty$ similar to the non-Markovian damping term.

\section{Effect of an anisotropy field on precession dynamics}
\label{sec:anis}
The theoretical analysis given in the previous section can be applied to the case with $H_{k}\neq 0$ by replacing $\xi$ with $\xi_{k}= \gamma\left(H+H_{k}m_{z}\right)\tau_{c}$. Following the same procedure as for $H_{k}=0$ Eq. \eqref{eq:gL} can be expressed as
\begin{align}
  \label{eq:renormalizedLLG_effective_Hk_2}
  \dot{\bm{m}}
   & =
  -\gamma \bm{m}\times
  \Bigl(
  H
  +
  \frac{\alpha\,H\,\xi_{k}}{1+\xi_{k}^{2}}
  m_{z}\
  +
  \frac{\alpha\,H_{k}\,\xi_{k}}{1+\xi_{k}^{2}}
  m_{z}^{2}
  \Bigr)\bm{e}_{z}
  \nonumber \\
   &
  -\gamma \bm{m}\times\bm{r}
  + \frac{\alpha}{1+\xi_{k}^{2}}\,\mathbf{m}\times\dot{\mathbf{m}}
  + O(\alpha^{2}).
\end{align}
The second and the third terms in the bracket can be regarded as the fictitious uniaxial and unidirectional anisotropy fields caused by the non-Markovian damping. Similar to the results for $H_{k}=0$ the precession angular velocity is maximized at $\xi_{k}=1$. The renormalized damping constant is given by $\alpha/(1 +\xi_{k}^{2})$ which is a monotonic decreasing function of $\xi_{k}$ and vanishes in the limit of $\xi_{k}\to \infty$.

\section{A possible experiment to determine the correlation time}
\label{sec:possible}
Based on the results shown in Secs. \ref{sec:hkzero} and \ref{sec:anis} we propose a possible experiment to determine the correlation time, $\tau_{c}$. Similar to the previous sections we first discuss the case without anisotropy field, i.e. $H_{k} = 0$, and then extend the discussion to the case with $H_{k}\neq 0$.

\begin{figure}[t]
  \centerline{\includegraphics [width=\columnwidth] {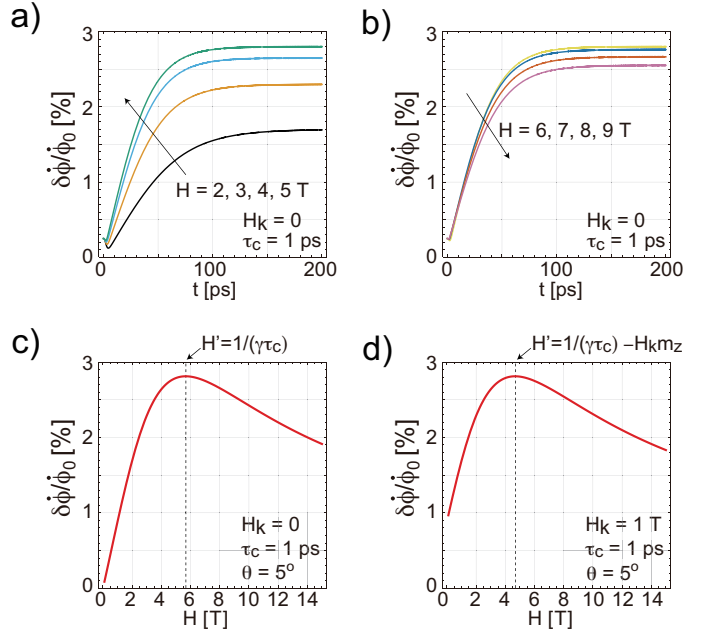}}
  \caption{
  \label{fig3}
  (a) $\tau_{c}$ dependence of $\delta\dot{\phi}/\dot{\phi}_{0}$ at $\theta=5^{\circ}$. From top to bottom the external field is $H=2, 3, 4, 5$ T. The parameters are $H_{k}=0$, and $\tau_{c} = 1$ ps.
  (b) The same plot as panel (a) for $H \ge 5$ T. From top to bottom the external field is $H=6, 7, 8, 9$ T.
  (c) $H$ dependence of $\delta\dot{\phi}/\dot{\phi}_{0}$ at $\theta=5^{\circ}$ obtained by solving  Eqs. \eqref{eq:gL_diff1} and \eqref{eq:gL_diff2}.
  The parameters are $H_{k}=0$, and $\tau_{c} = 1$ ps. The critical value of the external field, $H^{\prime}=1/(\gamma \tau_{c})$, is indicated by the thin vertical dotted line.
  (d) The same plot as panel (c) for $H_{k}=1$ T. The thin vertical dotted line indicates the critical value of the external field, $H^{\prime}=1/(\gamma\tau_{c})-H_{k}m_{z}$.
  }
\end{figure}

In Figs. \ref{fig3}(a) and \ref{fig3}(b) we show the temporal evolution of the enhancement of angular velocity, $\delta\dot{\phi}/\dot{\phi}_{0}$, obtained by the solving  Eqs. \eqref{eq:gL_diff1} and \eqref{eq:gL_diff2} for various values of $H$. The increment of the precession angular velocity is defined as $\delta \dot{\phi}= \dot{\phi}-\dot{\phi}_{0}$. The initial state and the correlation time are assumed to be $\bm{m}=(1,0,0)$ and $\tau_{c}=1$ ps, respectively. As shown in Fig. \ref{fig3}(a), $\delta\dot{\phi}/\dot{\phi}_{0}$ increases with increase of $H$ for $H \le 5 $T. Once the external field exceeds the critical value of $1/(\gamma\tau_{c}) = 5.7$ T, $\delta\dot{\phi}/\dot{\phi}_{0}$ decreases with increase of $H$ as shown in Fig. \ref{fig3} (b). The results suggest that correlation time can be determined by analyzing the external field that maximizes the enhancement of the precession angular velocity.

Figure \ref{fig3}(c) shows the $H$ dependence of $\delta\dot{\phi}/\dot{\phi}_{0}$ at $\theta=5^{\circ}$ where $\delta\dot{\phi}/\dot{\phi}_{0}$ is almost saturated. The enhancement is maximized at the critical value of the external field, $H^{\prime}=5.7$ T. The correlation time is calculated as $\tau_{c}=1/(\gamma H^{\prime}) = 1$ ps.

If the system has a uniaxial anisotropy field, $H_{k}$, the enhancement of the precession angular velocity is maximized at $H^{\prime}=1/(\gamma\tau_{c})-H_{k}m_{z}$ as shown in Fig. \ref{fig3}(d). The correlation time is obtained as $\tau_{c} = 1 / \gamma(H^{\prime} + H_{k} m_{z})$.

The above analysis is expected to be performed experimentally using the time resolved magneto optical Kerr effect measurement technique. In the practical experiments the analysis can be simplified as follows. The polar angle of the initial state is not necessarily large. It can be small as far as the precession angular velocity can be measured. Instead of analyzing $\delta\dot{\phi}/\dot{\phi}_{0}$, one can analyze $\dot{\phi}/H$ or $\dot{\phi}/(H+H_{k}m_{z})$ because they are maximized at the same value of $H$ as $\delta\dot{\phi}/\dot{\phi}_{0}$. Since the required magnetic field is as high as 10 T, a superconducting magnet \cite{weijers2010} is required.

\section{Summary}
\label{sec:summary}
In summary we theoretically analyze the ultrafast precession dynamics of a small magnet with non-Markovian damping.  Assuming $\alpha \ll 1$, we derive the effective LLG equation valid for any values of $\tau_{c}$, which is a direct extension of Miyazaki and Seki's work\cite{miyazaki1998}. The derived effective LLG equation reveals the condition for maximizing $\dot{\phi}$ in terms of $H$ and $\tau_{c}$. Based on the results we propose a possible experiment for determination of $\tau_{c}$, where $\tau_{c}$ can be determined from the external field that maximizes $\delta\dot{\phi}/\dot{\phi}_{0}$.



\end{document}